\documentclass[amssymb,aps,nofootinbib,twocolumn,superscriptaddress,showpacs]{revtex4} 
\usepackage[]{graphicx}
\usepackage[]{amsmath}
\usepackage{hyperref}

\def\ket#1{| #1 \rangle}
\def\bra#1{\langle #1 |}
\def\bracket#1#2{\langle #1 | #2 \rangle}
\def\kb#1#2{| #1 \rangle\!\langle #2 |}

\def\cD{\mathcal{D}}

\def\cL{\mathcal{L}}

\def\cO{\mathcal{O}}

\begin{document}

\title{Preparing ground states of quantum many-body systems on a quantum computer}
\author{David Poulin\footnote{To whom correspondence should be addressed. Email: David.Poulin@USherbrooke.ca}}
\affiliation{D\'epartement de Physique, Universit\'e de Sherbrooke, QC, Canada}
\author{Pawel Wocjan}
\affiliation{School of Electrical Engineering and Computer Science, University of Central Florida, FL, USA}

\date{\today}

\begin{abstract}
Preparing the ground state of a system of interacting classical particles is an NP-hard problem. Thus, there is in general no better algorithm to solve this problem than exhaustively going through all $N$ configurations of the system to determine the one with lowest energy, requiring a running time proportional to $N$. A quantum computer, if it could be built, could solve this problem in time $\sqrt N$. Here, we present a powerful extension of this result to the case of interacting {\em quantum} particles, demonstrating that a quantum computer can prepare the ground state of a quantum system as efficiently as it does for classical systems. 
\end{abstract}


\maketitle

The simulation of quantum many-body systems is a notoriously hard problem in condensed matter physics, but it could easily be handled by a quantum computer \cite{Llo96b}. There is however one catch: while a quantum computer can naturally implement the dynamics of a quantum system --- i.e. solve Schr\"odinger's equation --- there was until now no general method to initialize the computer in a physically relevant state of the simulated system. 

For most physical applications, we are interested in the low-energy eigenstates of the Hamiltonian $H$ because they describe the most interesting phases of matter, e.g. ferromagnetism, superconductivity, quantum Hall effect, and Bose-Einstein condensation to name a few.  Unfortunately, preparing low-energy states is already a very difficult task even when $H$ describes a classical system. 

Indeed, this problem is an archetype of the complexity class NP. This class contains all decision problems, i.e. problems of the form ``Does $x$ satisfy the property $\cL$?", such that when the answer is ``yes", there exists a witness $w$ that can be used to prove this answer efficiently. More precisely, for each $x$ there exists a polynomial-size verification circuits $V_x$ such that 1) when $x\in \cL$ there exists a witness $w$ that will cause $V_x$ to output $1$, and 2) when $x \notin \cL$, all witnesses cause $V_x$ to output $0$. 

Consider for instance a local Ising model $H(\{\sigma\}) = \sum_{i<j} J_{ij} \sigma_i\sigma_j + \sum_i h_i\sigma_i$ on $n$ spins $\sigma_i \in \{0,1\}$. ``Is there a spin configuration $\sigma$ of energy less than $E$?" is a problem in NP. Indeed, when the answer is yes, the configuration $\sigma$ that achieves this low energy can serve as a witness. Verifying the answer boils down to computing the energy, which requires at most $n^2$ operations. Finally, when the answer is ``no", there is no configuration that can cause the verification procedure to accept. Clearly, an algorithm that solves this problem can be used to determine the ground state energy with little overhead. 

The Ising problem is in fact NP-complete, meaning that it is the hardest problem in the class \cite{B82a}. Even more surprising is the fact that approximating the energy of the system with an error that increases with the system size $n$ is just as hard as the exact case --- it is also NP-complete. This is a consequence of  a famous theorem on probabilistically checkable proofs (PCP) \cite{AS98a,ALMS98a}. Although some special cases can be solved efficiently \cite{BMRU80a,BMRU82a,HR01a}, there is in general no better algorithm to solve the Ising problem than systematically going through all $N=2^n$ spin configurations to determine the one with lowest energy.

Additionally to the immediate physical context, finding ground states provides a very natural setting for studying combinatorial optimization problems --- problems that consist in minimizing an objective function $H$ (playing the role of energy) over some configuration space. Optimization problems play a vital role in almost every branch of science, from computer science to statistical physics and computational biology \cite{CCPS98a}.  Determining a solution by exhaustive search is, in general, computationally prohibitive because the size $N$ of the search space grows exponentially with the input size. Given the practial importance of optimization problems, more efficient methods are highly desirable.

A common strategy to solve optimization problems is simulated annealing \cite{KGV83a}. Like its name suggests, this method imitates the process undergone by a metal that is heated to a high temperature and then slowly cooled to its configuration of lowest energy. If the cooling process is too fast, the system can become trapped in a local minimum, resulting in a failure of the algorithm. When the cooling is sufficiently slow however, thermal fluctuations should prevent this phenomenon from occurring. Thus, simulated annealing requires a detailed knowledge of the energy landscape and therefore cannot be applied to all minimization problems. It was shown recently that a simulated annealing algorithm operated on a quantum computer achieves a quadratic speed-up over classical annealing \cite{SBBK08a}. Whether the method can minimize the energy of a quantum system as efficiently is unknown.

Adiabatic quantum computation is another method to tackle this class of problems with a quantum computer. The adiabatic theorem asserts that a system prepared in the instantaneous ground state of a Hamiltonian that varies slowly in time will remain in the ground state. The ground state of $H$ can thus be prepared by choosing a time-dependent Hamiltonian with a simple initial ground state and slowly changing it to $H$.  This algorithm was applied to randomly generated instances of an NP-complete problem \cite{FGGL01a}. The algorithm  worked well for the small examples that could be simulated on a classical computer. It was later shown  \cite{vDV08a} however that the particular interpolation scheme suggested \cite{FGGL01a} fails for satisfiability problems, and the best known upper bound is a running time of $N$ \cite{vDV08a} (poly-logarithmic corrections are ignored throughout).  In principle, this technique can also be applied to minimize the energy of a quantum Hamiltonian, but little is known about its performances in that case.  

Finally, Grover's algorithm \cite{Gro96a} can find the ground state of a classical system in $\sqrt N$ steps. Given a projector $R$ and a state $\psi$ with $\|R\ket\psi\|^2 = q >0$, Grover's algorithm consists of a sequence of two reflections, $I-2R$ and $I-2\kb\psi\psi$. Repeating this sequence $1/\sqrt q$ times has the effect of projecting $\psi$ onto the image of $R$, with small corrections that can safely be ignored for the present discussion. Note that the value of $q$ must be approximately known, and this can be achieved by quantum counting \cite{BHT98b}. Choosing $R$ to be the projector on $H < E$ and $\psi$ a uniform superposition of all spin configurations yields,  after at most $\sqrt N$ iterations, a state of energy less than $E$. The ground state is obtained by ``sweeping" the value of $E$.  Although this remains an exponential scaling, it is significantly faster than a brute force search, and there are indications that this scaling is optimal \cite{BBBV97a}.

At first sight, it seems like this last technique could be used to find the ground state of a quantum many-body system just as well. All that is needed is a method to implement a projector $R$ onto the low-energy states of the system, i.e. $H<E$ for some given $E$. Combining this method with Grover's algorithm on an initial random state would create the desired outcome with high probability. In fact, it is not necessary to initialize the system in a truly random state, but instead it can be randomly selected among all {\em stabilizer} states. These have all the essential properties of random states and, most importantly, can be prepared with at most $n^2$ operations \cite{DCEL06a}. Unfortunately, there is no known procedure to implement the projector on $H<E$. The phase estimation algorithm \cite{Kit95a} comes close however: it can be used to estimate the energy of any local Hamiltonian with a polynomial small error and failure probability.

To describe this algorithm, it is convenient to assume that $H$ has been normalized such that $\|H\| < 1$ and to consider its spectral decomposition, $H\ket a = \varphi_a \ket a$. The phase estimation algorithm uses $k$ auxiliary qubits initially in the state 0 and, given an eigenstate $\ket a$ of $H$, produces the output $\ket a \otimes \ket{\varphi_a}$ where  
\begin{equation}
\ket{\varphi_a} = \frac{1}{\sqrt{2^k}} \sum_j e^{-i2\pi \varphi_a j} \ket j.
\label{eq:momentum}
\end{equation}
These are ``momentum" states, so the value of $\varphi_a$ can be estimated via Fourier transform. Hence, we can implement an approximation $R$ of the projector on $H<E$ by running the phase estimation algorithm and projecting the auxiliary qubits onto the subspace of low momentum. Combining this method with Grover's algorithm should thus yield a good approximation of the ground state. 

However, a detailed analysis (see Appendix \ref{naive}) of this ``naive" approach reveals a failure. The problem is that the projector $R$ constructed from phase estimation is only an approximation of $H<E$ and errors can build up during the amplification procedure. There are two sources of errors. Firstly, the quantum computer cannot exactly reproduce the dynamics of the many-body system. This is not a problem however since a $1/{\rm poly}(n)$ accuracy can be achieved using a Trotter-Suzuki decomposition at a polynomial cost \cite{Llo96b}, and this error does not build up. We will henceforth safely ignore this source of error. 

Secondly, even when the energy $\varphi_a$ associated with $\ket a$ is well above the acceptance threshold $E$, there is a small probability that phase estimation will diagnose it as being smaller than $E$. It is these imperfections that cause the algorithm to fail because they build up during amplification. The typical outcome of this naive algorithm is an entangled state of the system and auxiliary qubits  rather than a low-energy state of the system qubits tensored with the all-zero state of the auxiliary qubits. Detailed knowledge of the energy landscape --- such as the presence of an energy gap --- could be used to circumvent this effect, but in general the method will fail.

We will now present our algorithm that works for all local Hamiltonians. We proceed by making two modifications to the naive algorithm. A detailed analysis is presented in the appendices. The first modification is to run the algorithm backward: we initialize the system qubits in a random state $\ket\psi = \sum_a \alpha_a \ket{a}$, the auxiliary qubits in a low momentum state $\ket\mu$ (c.f. Eq.~\ref{eq:momentum}), and execute the inverse of the phase estimation algorithm. This produces the state
\begin{equation}
\ket\Phi = \sum_a \alpha_a  \bracket{\varphi_a}{\mu} \ket a \otimes  \ket{0_k} + \ldots
\label{eq:Phi}
\end{equation}
where the ellipsis represents terms where the auxiliary qubits are in a state orthogonal to $\ket{0_k}$. The factor $|\bracket{\varphi_a}{\mu}|$ is a function of $\mu-\varphi_a$ peaked at 0 with a width $2^{-k}$. Thus, we can use Grover's algorithm to amplify the all-zero state of the auxiliary qubits and obtain a state that is mostly a superposition of those eigenstates of $H$ with eigenvalues close to $\mu$, i.e. the amplitude of each term in the superposition gets re-weighted by $|\bracket{\varphi_a}{\mu}|$. This procedure truly acts as a filter, suppressing the amplitude of eigenstates outside its bandwidth for benefit of the eigenstates inside the bandwidth. Moreover, the auxiliary qubits are systematically returned to $0$ as desired. 

Unfortunately, this is still not sufficient for our purpose because the filter has a heavy tail. There is an exponential number of states with energy outside the bandwidth, so unless their amplitude is exponentially suppressed, they can significantly shift the energy of the state. The filter we have constructed offers a polynomial suppression; we need a filter that drops more abruptly outside its bandwidth.

This requires a second modification to the naive algorithm and is realized by repeating the phase estimation $\eta$ times, using a total of $\eta k$ auxiliary qubits. We obtain the same state $\Phi$ as above (c.f. Eq.~\ref{eq:Phi}), except that the factor $\bracket{\varphi_a}{\mu}$ is now raised to the $\eta$th power.  For those $\varphi_a$ that are within $2^{-k}/\sqrt \eta$ of $\mu$, this factor is at least $1/2$. Thus, the overlap of this state with the projector $Q = I_n \otimes\kb{0_k}{0_k}^{\otimes \eta}$ will typically be $\|Q\ket\Phi\|^2 \geq \frac{m}{2N}$ where $m$ is the number of eigenstates of $H$ in the bandwidth of the filter. 

Thus, Grover's algorithm can be used to amplify this overlap to nearly 1 in a time at most $\sqrt {N}$. When the overlap of the state with the filter is too small, i.e. if $\|Q\ket\Phi\|^2 < 1/N^2$ say, this step will fail and the algorithm will abort.  Choosing $k \sim \log_2(\frac 1\epsilon)$ and $\eta \sim n$ yields, after a successful application of Grover's amplification, a state of energy $\mu\pm\epsilon$ as desired (see Appendix \ref{ground}). To summarize, this algorithm acts as a filter on the energy. The position $\mu$ and width $\epsilon \sim 1/{\rm poly}(n)$ of the filter are specified by the user. When no eigenstates of $H$ lie within the filter's bandwidth, the algorithm aborts as desired.

Note that the method can be adapted in a straightforward way to produce thermal distributions of the system at any temperature $T \geq 1/{\rm poly}(n)$. We could in a first step combine our method with quantum counting \cite{BHT98b} to estimate the density of states $\cD(E) = \sum_a \delta(\varphi_a-E) $  with a $1/{\rm poly}(n)$ resolution. We could then choose an energy scale $E$ at random according to the distribution $P(E) \sim e^{E/k_BT} \cD(E)$ and use our algorithm to prepare a state of energy close to $E$. 

Analogously to the classical case, determining the ground state energy of a local quantum many-body system within accuracy $1/{\rm poly}(n)$ is a complete problem for the complexity class known as Quantum Merlin and Artur (QMA) \cite{KSV02a}. Whether the problem remains complete when an extensive error is tolerated is unknown, but would be a natural quantum extension of the PCP theorem. Indeed, QMA is a natural generalization of NP: it is defined similarly except that both the witness and the verification circuit $V_x$ are quantum mechanical. Beside the $n$ witness qubits on which it operates, $V_x$ can also make use of $h\sim{\rm poly}(n)$  auxiliary qubits initialized in the state $0$ that serve as a scratchpad during the computation. The output of the verification procedure is obtained by measuring the first output qubit of the circuit. Because of the intrinsic randomness of quantum mechanics, this procedure is probabilistic:  1) when $x\in \cL$, there exists a witness $w$ that will cause $V_x$ to output $1$ with probability greater than $u$, and 2) when $x \notin \cL$, all witnesses cause $V_x$ to output $1$ with probability less than $v$ where $u-v > 1/{\rm poly}(n)$. 

The completeness of the local Hamiltonian problem for the class QMA  suggests that our algorithm can be used to solve all these problems and prepare the relevant witness in a time $\sqrt{2^n}$. This is not quite right because the mapping to the local Hamiltonian problem does not preserve the size of the witness. Nevertheless, a small modification almost does the trick.

A good witness is a $n+h$ qubit state $\psi$ with 1) all $h$ ancillary qubits in the state $0$ --- summarized by $Q\ket\psi = \ket\psi$ where $Q$ is the projector onto the all-zero state of the auxiliary qubits --- and 2) a probability at least $v$ of outputting $1$ at the verification procedure --- summarized by $\|R\ket\psi\|^2\geq v$ where $R$ is the projector associated with the verification procedure. Thus, the tasks of preparing a good witness boils down to producing a state that maximizes the overlap with two projectors $Q$ and $R$. When $[Q,R] = 0$, this task can be accomplished straightforwardly with Grover's algorithm, but additional efforts are required in the general case. Indeed, if we start say with a random state in the image of $Q$ and amplify the projector $R$, we will typically obtain a state that is mostly supported outside the image of $Q$, unless $[Q,R]=0$. 

\begin{figure}[t!]
\includegraphics{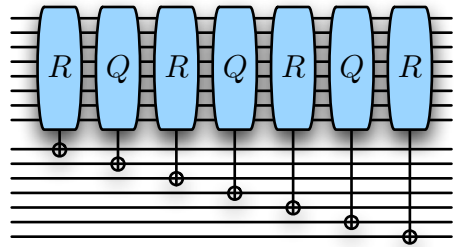}
\caption{The circuit consists of a sequence of measurements of $R$ and $Q$. The results are coherently imprinted on $k$ auxiliary qubits.}
\label{fig:circuit}
\end{figure}

The method we propose to solve this problem is a modification of our algorithm that builds on the work of Mariott and Watrous \cite{MW05a}. The main modification is to replace the phase estimation algorithm by the circuit of Figure \ref{fig:circuit}. It consists of a sequence of $k$ (odd) alternating measurements of $R$ and $Q$ whose outcomes are coherently recorded on $k$ auxiliary qubits initially in the state 0. The behavior of this circuit on an eigenstate $\ket a$ of the operator $QRQ$ with eigenvalue $p_a$ can be analyzed using a result of Jordan \cite{J75a} (see Appendix \ref{Jordan}). The state of the $k$ auxiliary qubits becomes a superposition of all sequences of $0$ and $1$, and the amplitude of consecutive distinct outcomes, i.e. the amplitude associated to each ``switches" from 0 to 1 or vice versa, is $\sqrt{1-p_a}$. Thus, counting the number of switches in the measurement outcomes allows us to estimate the eigenvalue $p_a$ of the state.

The situation is therefore analogous to phase estimation, except that the eigenvalue $p_a$ is not encoded in a momentum state but in a state with a certain number of switches between the outcomes 0 and 1. Accordingly, we must replace the momentum state used in our algorithm by a state with the right distribution of switches $\ket{\mu} = \sum_{j \in\{0,1\}^k} (\sqrt{\mu})^{k-s(j)} (\sqrt{1-\mu})^{s(j)} (-1)^{\ell(j)} \ket{j}$ where $s$ denotes the number of switches and $\ell$ the number of pairs of consecutive $0$'s.  The bandwidth $\epsilon$ is adjusted by setting $k = \frac{2\mu(1-\mu)}{\epsilon ^2}$. One important advantage of this type of filter state is that it drops very abruptly outside its bandwidth, $|\bracket{\mu}{p}|$ is essentially proportional to a normal distribution centered at $p = \mu$ and of variance $2\mu(1-\mu)/k^2$. Thus, there is no need for multiple copies of the filter state and the rest of the algorithm proceeds as before. 

This more general algorithm does not perform as well as the algorithm used for local Hamiltonians because it searches over a larger Hilbert space: the space of the witness and the scratchpad. This is to be expected since it makes no assumption about the structure of the verification procedure $V_x$.  Note however that all known ``natural" problems in QMA --- e.g. non-identity check \cite{JWB05a}, consistency of quantum states \cite{L06a}, $N$-representability \cite{LCV07a}, and  zero-error capacity of quantum channels \cite{BS07a} --- use a scratchpad of only logarithmic size, so in those cases the running time is the same as for local Hamiltonians. It is tempting to conjecture that the scratchpad of all problems in QMA can be reduced to this size. 

To summarize, we have presented a method to prepare ground and thermal states of quantum many-body systems on a quantum computer. The time required by our algorithm is equal to the square-root of the Hilbert space dimension of the system --- the same time required to prepare the ground state of a classical many-body system. This represents a speed up by a power of 6 compared to exact diagonalization, which in general is the only available technique to accomplish this task on a classical computer.  It is perhaps surprising that this task cannot be accomplished by a straightforward combination of phase estimation and Grover's algorithm, but our analysis of this strategy reveals an important failure and more elaborate methods were required. A quantum computer, if it could be built, could serve as an efficient simulator of quantum many-body systems. The method we have presented would complement this simulation by initializing the computer in a low-energy state of the simulated system.

 With some modifications, our algorithm can be used to solve and prepare relevant witnesses of all problems in the complexity class QMA, the quantum generalization of NP. In that case, the physical task consists of preparing a state that has a large overlap with two projectors. Problems in NP form a special case where those projectors commute and can be solved straightforwardly using Grover's algorithm. However, in the general case the projectors do not commute and more sophisticated techniques were required.

We acknowledge Andrew Childs, Peter H\o yer, Dominik Janzing, John Preskill, and John Watrous for stimulating discussions. This work began while P.W. was visiting D.P. who was a postdoctoral scholar at Caltech.


\begin{thebibliography}{24}
\expandafter\ifx\csname natexlab\endcsname\relax\def\natexlab#1{#1}\fi
\expandafter\ifx\csname bibnamefont\endcsname\relax
  \def\bibnamefont#1{#1}\fi
\expandafter\ifx\csname bibfnamefont\endcsname\relax
  \def\bibfnamefont#1{#1}\fi
\expandafter\ifx\csname citenamefont\endcsname\relax
  \def\citenamefont#1{#1}\fi
\expandafter\ifx\csname url\endcsname\relax
  \def\url#1{\texttt{#1}}\fi
\expandafter\ifx\csname urlprefix\endcsname\relax\def\urlprefix{URL }\fi
\providecommand{\bibinfo}[2]{#2}
\providecommand{\eprint}[2][]{\url{#2}}

\bibitem[{\citenamefont{Lloyd}(1996)}]{Llo96b}
\bibinfo{author}{\bibfnamefont{S.}~\bibnamefont{Lloyd}},
  \bibinfo{journal}{Science} \textbf{\bibinfo{volume}{273}},
  \bibinfo{pages}{1073} (\bibinfo{year}{1996}).

\bibitem[{\citenamefont{Barahona}(1982)}]{B82a}
\bibinfo{author}{\bibfnamefont{F.}~\bibnamefont{Barahona}},
  \bibinfo{journal}{J. Phys. A. Math. Gen.} \textbf{\bibinfo{volume}{15}},
  \bibinfo{pages}{3241} (\bibinfo{year}{1982}).


\bibitem[{\citenamefont{Arora and Safra}(1998)}]{AS98a}
\bibinfo{author}{\bibfnamefont{S.}~\bibnamefont{Arora}} \bibnamefont{and}
  \bibinfo{author}{\bibfnamefont{S.}~\bibnamefont{Safra}}, \bibinfo{journal}{J.
  ACM} \textbf{\bibinfo{volume}{45}}, \bibinfo{pages}{70}
  (\bibinfo{year}{1998}).

\bibitem[{\citenamefont{Arora et~al.}(1998)\citenamefont{Arora, Lund, Motwani,
  and Szegedy}}]{ALMS98a}
\bibinfo{author}{\bibfnamefont{S.}~\bibnamefont{Arora}},
  \bibinfo{author}{\bibfnamefont{C.}~\bibnamefont{Lund}},
  \bibinfo{author}{\bibfnamefont{R.}~\bibnamefont{Motwani}}, \bibnamefont{and}
  \bibinfo{author}{\bibfnamefont{M.}~\bibnamefont{Szegedy}},
  \bibinfo{journal}{J. ACM} \textbf{\bibinfo{volume}{45}}, \bibinfo{pages}{501}
  (\bibinfo{year}{1998}).

\bibitem[{\citenamefont{Bieche et~al.}(1980)\citenamefont{Bieche, Maynard,
  Rammal, and Uhry}}]{BMRU80a}
\bibinfo{author}{\bibfnamefont{I.}~\bibnamefont{Bieche}},
  \bibinfo{author}{\bibfnamefont{R.}~\bibnamefont{Maynard}},
  \bibinfo{author}{\bibfnamefont{R.}~\bibnamefont{Rammal}}, \bibnamefont{and}
  \bibinfo{author}{\bibfnamefont{J.}~\bibnamefont{Uhry}}, \bibinfo{journal}{J.
  Phys. A} \textbf{\bibinfo{volume}{13}}, \bibinfo{pages}{2553}
  (\bibinfo{year}{1980}).

\bibitem[{\citenamefont{Barahona et~al.}(1982)\citenamefont{Barahona, Maynard,
  Rammal, and Uhry}}]{BMRU82a}
\bibinfo{author}{\bibfnamefont{F.}~\bibnamefont{Barahona}},
  \bibinfo{author}{\bibfnamefont{R.}~\bibnamefont{Maynard}},
  \bibinfo{author}{\bibfnamefont{R.}~\bibnamefont{Rammal}}, \bibnamefont{and}
  \bibinfo{author}{\bibfnamefont{J.}~\bibnamefont{Uhry}}, \bibinfo{journal}{J.
  Phys. A} \textbf{\bibinfo{volume}{15}}, \bibinfo{pages}{673}
  (\bibinfo{year}{1982}).

\bibitem[{\citenamefont{Hartmann and Rieger}(2001)}]{HR01a}
\bibinfo{author}{\bibfnamefont{A.}~\bibnamefont{Hartmann}} \bibnamefont{and}
  \bibinfo{author}{\bibfnamefont{H.}~\bibnamefont{Rieger}},
  \emph{\bibinfo{title}{Optimization algorithms in physics}}
  (\bibinfo{publisher}{Wiley-VCH}, \bibinfo{address}{Berlin},
  \bibinfo{year}{2001}).

\bibitem[{\citenamefont{Cook et~al.}(1998)\citenamefont{Cook, Cunninham,
  Pulleyblank, and Schrijver}}]{CCPS98a}
\bibinfo{author}{\bibfnamefont{W.}~\bibnamefont{Cook}},
  \bibinfo{author}{\bibfnamefont{W.}~\bibnamefont{Cunninham}},
  \bibinfo{author}{\bibfnamefont{W.}~\bibnamefont{Pulleyblank}},
  \bibnamefont{and}
  \bibinfo{author}{\bibfnamefont{A.}~\bibnamefont{Schrijver}},
  \emph{\bibinfo{title}{Combinatorial Optimization}} (\bibinfo{publisher}{J.
  Wiley and Sons}, \bibinfo{year}{1998}).

\bibitem[{\citenamefont{Kirkpatrick et~al.}(1983)\citenamefont{Kirkpatrick,
  Gelatt, and Vecchi}}]{KGV83a}
\bibinfo{author}{\bibfnamefont{S.}~\bibnamefont{Kirkpatrick}},
  \bibinfo{author}{\bibfnamefont{C.~J.} \bibnamefont{Gelatt}},
  \bibnamefont{and} \bibinfo{author}{\bibfnamefont{M.}~\bibnamefont{Vecchi}},
  \bibinfo{journal}{Science} \textbf{\bibinfo{volume}{220}},
  \bibinfo{pages}{4598} (\bibinfo{year}{1983}).

\bibitem[{\citenamefont{Somma et~al.}(2008)\citenamefont{Somma, Boixo, Barnum,
  and Knill}}]{SBBK08a}
\bibinfo{author}{\bibfnamefont{R.}~\bibnamefont{Somma}},
  \bibinfo{author}{\bibfnamefont{S.}~\bibnamefont{Boixo}},
  \bibinfo{author}{\bibfnamefont{H.}~\bibnamefont{Barnum}}, \bibnamefont{and}
  \bibinfo{author}{\bibfnamefont{E.}~\bibnamefont{Knill}},
  \emph{\bibinfo{title}{Quantum simulations of classical annealing processes}}
  (\bibinfo{year}{2008}), \eprint{arXiv:0804.1571}.

\bibitem[{\citenamefont{Farhi et~al.}(2001)\citenamefont{Farhi, Goldstone,
  Gutmann, Lapan, Lundgren, and Preda}}]{FGGL01a}
\bibinfo{author}{\bibfnamefont{E.}~\bibnamefont{Farhi}},
  \bibinfo{author}{\bibfnamefont{J.}~\bibnamefont{Goldstone}},
  \bibinfo{author}{\bibfnamefont{S.}~\bibnamefont{Gutmann}},
  \bibinfo{author}{\bibfnamefont{J.}~\bibnamefont{Lapan}},
  \bibinfo{author}{\bibfnamefont{A.}~\bibnamefont{Lundgren}}, \bibnamefont{and}
  \bibinfo{author}{\bibfnamefont{D.}~\bibnamefont{Preda}},
  \bibinfo{journal}{Science} \textbf{\bibinfo{volume}{292}},
  \bibinfo{pages}{472} (\bibinfo{year}{2001}).

\bibitem[{\citenamefont{{van Dam} and Varizani}(2008)}]{vDV08a}
\bibinfo{author}{\bibfnamefont{W.}~\bibnamefont{{van Dam}}} \bibnamefont{and}
  \bibinfo{author}{\bibfnamefont{U.}~\bibnamefont{Varizani}},
  \emph{\bibinfo{title}{Limits on quantum adiabatic optimization}},
  \bibinfo{howpublished}{premilinary draft} (\bibinfo{year}{2008}).
  
  \bibitem[{\citenamefont{Grover}(1996)}]{Gro96a}
\bibinfo{author}{\bibfnamefont{L.}~\bibnamefont{Grover}}, in
  \emph{\bibinfo{booktitle}{Proc. 28th Annual ACM Symposium on the Theory of
  Computation}} (\bibinfo{publisher}{ACM Press, New York},
  \bibinfo{address}{New York, NY}, \bibinfo{year}{1996}),
  \bibinfo{pages}{212}.

\bibitem[{\citenamefont{Brassard et~al.}(1998)\citenamefont{Brassard, Hoyer,
  and Tapp}}]{BHT98b}
\bibinfo{author}{\bibfnamefont{G.}~\bibnamefont{Brassard}},
  \bibinfo{author}{\bibfnamefont{P.}~\bibnamefont{Hoyer}}, \bibnamefont{and}
  \bibinfo{author}{\bibfnamefont{A.}~\bibnamefont{Tapp}}, in
  \emph{\bibinfo{booktitle}{Automata, Languages and Programming, Proceedings of
  ICALP'98}}, edited by \bibinfo{editor}{\bibfnamefont{K.~G.}
  \bibnamefont{Larsen}},
  \bibinfo{editor}{\bibfnamefont{S.}~\bibnamefont{Skyum}}, \bibnamefont{and}
  \bibinfo{editor}{\bibfnamefont{G.}~\bibnamefont{Winskel}}
  (\bibinfo{publisher}{Springer Verlag}, \bibinfo{address}{Berlin, Germany},
  \bibinfo{year}{1998}), vol. \bibinfo{volume}{1443} of
  \emph{\bibinfo{series}{Lecture Notes in Computer Science}}, 
  \bibinfo{pages}{820}.

\bibitem[{\citenamefont{Bennett et~al.}(1997)\citenamefont{Bennett, Bernstein,
  Brassard, and Vazirani}}]{BBBV97a}
\bibinfo{author}{\bibfnamefont{C.~H.} \bibnamefont{Bennett}},
  \bibinfo{author}{\bibfnamefont{E.}~\bibnamefont{Bernstein}},
  \bibinfo{author}{\bibfnamefont{G.}~\bibnamefont{Brassard}}, \bibnamefont{and}
  \bibinfo{author}{\bibfnamefont{U.}~\bibnamefont{Vazirani}},
  \bibinfo{journal}{SIAM J. Comput.} \textbf{\bibinfo{volume}{26}},
  \bibinfo{pages}{1510} (\bibinfo{year}{1997}).

\bibitem[{\citenamefont{Dankert et~al.}(2006)\citenamefont{Dankert, Cleve,
  Emerson, and Livine}}]{DCEL06a}
\bibinfo{author}{\bibfnamefont{C.}~\bibnamefont{Dankert}},
  \bibinfo{author}{\bibfnamefont{R.}~\bibnamefont{Cleve}},
  \bibinfo{author}{\bibfnamefont{J.}~\bibnamefont{Emerson}}, \bibnamefont{and}
  \bibinfo{author}{\bibfnamefont{E.}~\bibnamefont{Livine}},
  \emph{\bibinfo{title}{Exact and approximate 2-designs: construction and
  application}} (\bibinfo{year}{2006}), \eprint{quant-ph/0606161}.

\bibitem[{\citenamefont{Kitaev}(1995)}]{Kit95a}
\bibinfo{author}{\bibfnamefont{A.}~\bibnamefont{Kitaev}},
  \emph{\bibinfo{title}{Quantum measurements and the Abelian stabilizer
  problem}} (\bibinfo{year}{1995}), \eprint{quant-ph/9511026}.

\bibitem[{\citenamefont{Kitaev et~al.}(2002)\citenamefont{Kitaev, Shen, and
  Vyalyi}}]{KSV02a}
\bibinfo{author}{\bibfnamefont{A.~Y.} \bibnamefont{Kitaev}},
  \bibinfo{author}{\bibfnamefont{A.~H.} \bibnamefont{Shen}}, \bibnamefont{and}
  \bibinfo{author}{\bibfnamefont{M.~N.} \bibnamefont{Vyalyi}},
  \emph{\bibinfo{title}{Classical and quantum computation}}, Graduate studies
  in mathematics (\bibinfo{publisher}{American Mathematical Society},
  \bibinfo{address}{Providence, Rhodes Island}, \bibinfo{year}{2002}).

\bibitem[{\citenamefont{Marriott and Watrous}(2005)}]{MW05a}
\bibinfo{author}{\bibfnamefont{C.}~\bibnamefont{Marriott}} \bibnamefont{and}
  \bibinfo{author}{\bibfnamefont{J.}~\bibnamefont{Watrous}},
  \bibinfo{journal}{Computational Complexity} \textbf{\bibinfo{volume}{14}},
  \bibinfo{pages}{122} (\bibinfo{year}{2005}).

\bibitem[{\citenamefont{Janzing et~al.}(2005)\citenamefont{Janzing, Wocjan, and
  Beth}}]{JWB05a}
\bibinfo{author}{\bibfnamefont{D.}~\bibnamefont{Janzing}},
  \bibinfo{author}{\bibfnamefont{P.}~\bibnamefont{Wocjan}}, \bibnamefont{and}
  \bibinfo{author}{\bibfnamefont{T.}~\bibnamefont{Beth}},
  \bibinfo{journal}{Int. J. of Quant. Info.} \textbf{\bibinfo{volume}{3}},
  \bibinfo{pages}{463} (\bibinfo{year}{2005}).

\bibitem[{\citenamefont{Liu}(2006)}]{L06a}
\bibinfo{author}{\bibfnamefont{Y.-K.} \bibnamefont{Liu}},
  \bibinfo{journal}{Proc. RANDOM} p. \bibinfo{pages}{438}
  (\bibinfo{year}{2006}).

\bibitem[{\citenamefont{Liu et~al.}(2007)\citenamefont{Liu, Christandl, and
  Verstraete}}]{LCV07a}
\bibinfo{author}{\bibfnamefont{Y.-K.} \bibnamefont{Liu}},
  \bibinfo{author}{\bibfnamefont{M.}~\bibnamefont{Christandl}},
  \bibnamefont{and}
  \bibinfo{author}{\bibfnamefont{F.}~\bibnamefont{Verstraete}},
  \bibinfo{journal}{Phys. Rev. Lett.} \textbf{\bibinfo{volume}{98}},
  \bibinfo{pages}{110503} (\bibinfo{year}{2007}).

\bibitem[{\citenamefont{Beigi and Shor}(2007)}]{BS07a}
\bibinfo{author}{\bibfnamefont{S.}~\bibnamefont{Beigi}} \bibnamefont{and}
  \bibinfo{author}{\bibfnamefont{P.}~\bibnamefont{Shor}},
  \emph{\bibinfo{title}{On the complexity of computing zero-error and {H}olevo
  capacity of quantum channels}} (\bibinfo{year}{2007}),
  \eprint{arXiv:0709.2090}.

\bibitem[{\citenamefont{Jordan}(1875)}]{J75a}
\bibinfo{author}{\bibfnamefont{C.}~\bibnamefont{Jordan}},
  \bibinfo{journal}{Bulletin de la S. M. F.} \textbf{\bibinfo{volume}{3}},
  \bibinfo{pages}{103} (\bibinfo{year}{1875}).

\end{thebibliography}

\appendix

\section{Jordan's result}
\label{Jordan}

Our main tool of analysis is a result attributed to Jordan \cite{J75a}, a more modern version of which can be found in \cite{MW05a}. The result states that any two projectors $Q$ and $R$ on an $N$-dimensional Hilbert space can be put in a simultaneous block-diagonal form, with blocks of size at most 2. Clearly, the operators $QRQ$ and $RQR$ share the same spectrum $\{p_a\}$, bounded between 0 and 1 (this can be seen by left and right polar decomposition of the operator $QR$). Their respective eigenvectors  $\{q_a\}$ and $\{r_a\}$ are in general different however. Since $Q$ and $R$ are purely contractive, it follows that if $p_a=1$ for some $a$, then the basis can be chosen such that $\ket{q_a} = \ket{r_a}$. When $p_a = 0$, then either $Q\ket{q_a} = 0$ or $R\ket{r_a} = 0$. For the intermediate values $0<p_a<1$, Jordan found that we can group the remaining eigenvectors in pairs that we denote $(q_a^0,q_a^1)$ and $(r_a^0,r_a^1)$ such that
\begin{align}
&\ket{q_a^0} = \sqrt{p_a} \ket{r_a^0} - \sqrt{1-p_a} \ket{r_a^1} \label{Jordan1}\\
&\ket{q_a^1} = \sqrt{1-p_a} \ket{r_a^0} + \sqrt{p_a} \ket{r_a^1} \\
&\ket{r_a^0} = \sqrt{p_a} \ket{q_a^0} + \sqrt{1-p_a} \ket{q_a^1} \\
&\ket{r_a^1} = \sqrt{1-p_a} \ket{q_a^0} - \sqrt{p_a} \ket{q_a^1},\label{Jordan4}
\end{align}
with the property that $Q\ket{q_a^b} = b \ket{q_a^b}$ and $R\ket{r_a^b} = b \ket{r_a^b}$. By enlarging the dimension of the Hilbert space (at most doubling its dimension) and allowing $p_i$ to take the values 0 and 1, we can assume that all eigenstates of $QRQ$ and $RQR$ are paired up in this fashion, so that both sets $\{q_a^b\}$ and $\{r_a^b\}$ are complete orthonormal bases.

\section{Failure of the naive algorithm}
\label{naive}

The naive algorithm to approximate the ground state of a local Hamiltonian consists of using phase estimation to mark the low-energy eigenstates of $H$, and then integrating this marking procedure in Grover's algorithm to produce a state of low energy. This strategy can easily be analyzed with the help of Jordan's result. Indeed, let $Q$ be the projector onto the all-zero state of the $k$ auxiliary qubits used by phase estimation and $R$ be the projector on the output of the phase estimation indicating an energy less than $E$. 

As mentioned in the main text, the unitary evolution $U = e^{-iH}$ cannot be realized perfectly in general, but this is a source of error that can be ignored because it doesn't build up during Grover's algorithm. Thus, we will assume here that the phase estimation algorithm implements $U$ exactly.

In this language, the state $q_a^1$ corresponds to a state of the form $\ket{a}\otimes\ket{0_k}$ where $\ket a$ is an eigenstate of $H$ with energy $\varphi_a$.  The eigenvalue $p_a$ of $QRQ$ associated to that state is equal to the probability that the phase estimation algorithm diagnoses $\ket a$ as having an energy less than $E$. This is where the finite success probability of the phase estimation algorithm comes into play. The important point is that even if the energy $\varphi_a$ is well below or well above $E$, there is a small probability that the algorithm will give the wrong answer. The effect is more pronounced when the actual energy of $\varphi_a$ is close to the acceptance threshold $E$, but it is the small error probabilities that will really cause a problem. The error is zero only when $\varphi_a$ can be expressed exactly in binary form with $k$ bits. Thus, unless very detailed knowledge of the spectrum of $H$ is available, these small errors are unavoidable.

Continuing with the naive algorithm, we prepare a random superposition $\ket\psi = \sum_a \alpha_a \ket{q_a^1}$, i.e. a random state of the $n$ system qubits with the $k$ auxiliary qubits in state 0, and use Grover's algorithm to amplify the image of $R$. This results in the state
\begin{eqnarray}
\ket{\psi'} &= &\frac{R\ket\psi}{\|R\ket\psi\|} \\
&=& \frac{\sum_a \alpha_a \sqrt{p_a} \ket{r_a^1}}{\sqrt{\sum_a |\alpha_a|^2 p_a}} \\
&=& \frac{\sum_a \alpha_a \left(\sqrt{p_a-p_a^2}\ket{q_a^0} - p_a \ket{q_a^1}\right)}{\sqrt{\sum_a |\alpha_a|^2 p_a}}.
\end{eqnarray}
We see that, as a consequence of the amplification of $R$, the state $\psi'$  is no longer supported only on the image of $Q$ as indicated by the presence of the $q_a^0$ in the state. The overlap with $Q$ is easily computed
\begin{eqnarray}
\bra{\psi'}Q\ket{\psi'} 
&=& \frac{\sum_a |\alpha_a|^2  p_a^2}{\sum_a |\alpha_a|^2 p_a}
\end{eqnarray}
and this can be arbitrarily small. For instance, if the vast majority of states have $p_a \sim 1/\sqrt N$ and a negligible fraction of them have $p_a$'s of order unity, this overlap is of order $1/\sqrt N$. 

This example summarizes the main problem of the naive approach that is overcomed by our method. We want to create a state that has a large overlap with two projectors $Q$ and $R$. When $[Q,R]=0$, we can simply prepare a random state in the image of $Q$ and use Grover's algorithm to amplify $R$. But when the two projectors do not commute, the amplification of $R$ can move the state almost completely outside the image of $Q$, so the technique fails. 

\section{Detailed analysis  for the ground state problem}
\label{ground}

We make use of an upper and a lower bound on the inner-product of two different momentum states (c.f. Eq.~\ref{eq:momentum}) used by our algorithm. 1) $|\bracket{\varphi}{\mu}| \leq \frac{1}{2^{k+1}|\varphi - \mu|}$ and 2) when $|\varphi-\mu|\le 2^{-k}/(2\pi\sqrt{\eta})$,  then $|\bracket{\varphi}{\mu}|^\eta \geq 1/2$. 

To prove the first bound, we use the inequality $|1-e^{ix}|\ge 2|x|/\pi$ for $-\pi\le x\le \pi$ and elementary algebra to obtain
\begin{eqnarray}
|\bracket{\varphi}{\mu}| 
& = & 
\frac{1}{2^k} 
\Bigg|  \sum_{j=0}^{2^k-1} e^{i 2\pi j (\varphi - \mu)} 
\Bigg|  \\
& = &
\frac{1}{2^k}
\Bigg| 
\frac{1-e^{i 2\pi 2^k (\varphi - \mu)}}{1-e^{i2\pi (\varphi - \mu)}}
\Bigg|  \\
& \le &
\frac{1}{2^{k-1}} 
\Big| 
\frac{1}{1-e^{i2\pi (\varphi - \mu)}}
\Big| \\
& \le &
\frac{1}{2^{k-1}} 
\Big|
\frac{\pi}{2\cdot 2\pi |\varphi - \mu|}
\Big| \\
& \le &
\frac{1}{2^{k+1}|\varphi - \mu|}
\end{eqnarray}
as claimed. 

For the second bound, assume that $|\varphi-\mu|\le 2^{-k}/(2\pi\sqrt{\eta})$.  We can use the fact that $|\bracket{\varphi}{\mu}|  \geq {\rm Re}(\bracket{\varphi}{\mu})$ to obtain
\begin{eqnarray}
|\bracket{\varphi}{\mu}| 
& = & 
\frac{1}{2^k} \Bigg|  \sum_{j=0}^{2^k-1} e^{i 2\pi j (\varphi - \mu)} \Bigg|  \\
& \ge &
\frac{1}{2^k} \Bigg|  \sum_{j=0}^{2^k-1} \cos(2\pi j (\varphi - \mu)) \Bigg|  \\
& \ge &
\frac{1}{2^k} \sum_{j=0}^{2^k-1} \cos(1/\sqrt{\eta}) \\
& = &
\cos(1/\sqrt{\eta}) \\
& \ge &
1-1/(2\eta)\,.
\end{eqnarray}
Raising to the $\eta$th power and using the inequality  $(1-x)^n > 1-nx$ for $n<1$ and $x>0$ gives the desired result. 

We will now demonstrate that our algorithm will work as advertised, i.e. produce a state of energy $\mu \pm\epsilon$, with high probability. We break this proof into two parts. First, we show that the algorithm does not systematically abort and second that when it does not abort it produces the right state. 

Remember that the algorithm aborts when the overlap $\|Q\ket\Phi\|^2 = \sum_a |\alpha_a|^2 |\bracket{\varphi_a}{\mu}|^2$ is less than $\cO(1/N)$. Using the lower bound derived above, we know that if there is at least one eigenvalue, say $\varphi_0$, that lies in the interval $\mu \pm 2^{-k}/(2\pi\sqrt\eta)$, this overlap will be at least $|\alpha_0|^2/2$. Since the state is chosen at random, the amplitude $|\alpha_0|$ has an exponentially high probability of being greater than $1/(2\sqrt N)$ which is sufficient for the success of the amplification procedure. Note that the interval  $\mu \pm 2^{-k}/(2\pi\sqrt\eta)$ has a polynomial width, i.e. $k$ scales only logarithmically with $n$ and $\eta$ polynomially. Thus, we can sweep all values of $\mu$ to this accuracy in polynomial time. 

We must now demonstrate that when the amplification procedure succeeds, then with very high probability it generates a state with the desired energy. We can assume in this case that $\|Q\ket\Phi\|> 1/N$ since otherwise the amplification procedure would have had an exponentially small chance of succeeding. Let us first upper bound the average energy:
\begin{eqnarray}
\langle H \rangle &=&  \frac{\bra\Phi Q H Q \ket\Phi}
{\|Q\ket\Phi\|^2} \\
&=&  \frac{\sum_a |\alpha_a|^2 |\bracket{\varphi_a}{\mu}|^{2\eta} \varphi_a}
{\|Q\ket\Phi\|^2} \\
&\leq&  \mu + \frac\epsilon 2 + \frac{\sum_a' |\alpha_a|^2 |\bracket{\varphi_a}{\mu}|^{2\eta} \varphi_a}
{\|Q\ket\Phi\|^2}
\end{eqnarray}
where $\sum_a'$ is the sum restricted to values of $a$ such that $\varphi_a \geq \mu + \epsilon/2$. Using the lower bound on $\|Q\ket\Phi\|$ and the fact that $0\leq \varphi \leq 1$ we obtain  
\begin{eqnarray}
\langle H \rangle 
&\leq&  \big(\mu + \frac\epsilon 2\big) + N^2  |\bracket{\mu+\epsilon/2}{\mu}|^{2\eta}.
\end{eqnarray}
Consequently, the desired bound is obtained when $N^2  |\bracket{\mu+\epsilon/2}{\mu}|^{2\eta} \leq \epsilon/2$. Using our upper bound on the inner product $|\bracket{\mu+\epsilon/2}{\mu}| $ yields
\begin{equation}
\frac{N^2}{(2^k\epsilon)^{2\eta}} \leq \epsilon/2
\end{equation}
which is satisfied with $k\geq 2\log_2(\frac1\epsilon)$ and $\eta \geq 1 + (n+1)/\log_2(\frac 1\epsilon)$.

Similar arguments produce an upper bound on the average energy
\begin{eqnarray}
\langle H \rangle 
&=&  \frac{\sum_a |\alpha_a|^2 |\bracket{\varphi_a}{\mu}|^{2\eta} \varphi_a}
{\|Q\ket\Phi\|^2} \\
&\geq& \frac{ \Big(\|Q\ket\Phi\|^2 - \sum_a' |\alpha_a|^2 |\bracket{\varphi_a}{\mu}|^{2\eta}\Big) (\mu - \frac\epsilon 2)}
{\|Q\ket\Phi\|^2}\nonumber
\end{eqnarray}
where $\sum_a'$ is now the sum restricted to values of $a$ such that $\varphi_a \leq \mu - \epsilon/2$. Using the lower bound on $\|Q\ket\Phi\|$ and the upper bound on the inner product $|\bracket{\varphi_a}{\mu}|$ gives
\begin{equation}
\langle H \rangle  \geq \Big(1-\frac{N^2}{(2^k\epsilon)^{2\eta}}\Big)\big(\mu - \frac\epsilon 2\big)
\end{equation}
which is greater than $\mu - \epsilon$ provided that $\eta \geq 1+(n+\log_2\mu)/\log_2(\frac1\epsilon)$.

Note that with a $k$ growing polynomially with $\log(n)$ and $\eta$ growing polynomially with $n$, the error on the energy decreases as $1/{\rm poly}(n)$ and the total running time is $\sqrt{2^n}{\rm poly}(n)$ as desired. 

\section{Detailed analysis  for problems in QMA}

The analysis of the general algorithm for problems in QMA follows similar lines, but the derivation of the filter function is slightly more involved. The verification circuit $V_x$ acts on $m = n+h$ qubits: the $n$ qubits containing the witness and a $h$-qubit scratchpad.  We make use of Jordan's result with $Q$ being the projector onto the all-zero state of the scratchpad and $R = V_x (\kb 11 \otimes I_{m-1})V_x^\dagger$ is the projector onto the accepting subspace of the verification procedure. Remember that the verification procedure takes inputs of the form $\ket \psi = \ket w\otimes \ket{0_h}$ and accepts with probability $\|(\kb 11 \otimes I_{m-1})V_x \ket\psi\|^2 = \|QRQ\ket\psi\|^2$. Thus, the eigenstates $\ket{q_a^1}$ of  $QRQ$ correspond to witnesses with all-zero scratchpads with an accepting probability $p_a$.

Our algorithm makes use of $k$ additional auxiliary qubits, for a total of $n+h+k$. Armed Eqs~(\ref{Jordan1}-\ref{Jordan4}), we see that the circuit of Figure \ref{fig:circuit} applied to the state $\ket{q_a}\otimes \ket{0_k}$ produces the outcome
\begin{equation*}
\sum_{j \in\{0,1\}^k} (\sqrt{p_a})^{k-s(j)} (\sqrt{1-p_a})^{s(j)} (-1)^{\ell(j)} \ket{r^{j_k}_a} \otimes \ket{j}.
\end{equation*}
Indeed, we get factor of $\sqrt{1-p_a}$ each time the outcome switches from 0 to 1 or vice versa, and a factor of $\sqrt{p_a}$ otherwise. There is an additional factor of $-1$ for every pair of consecutive 0's.

Let us now analyse our algorithm. We initialize the first $m$ qubits in a random state $\psi$ in the support of $R$. This is done by setting the first qubit in the state $\ket 1$ and the remaining $m-1$ qubits in a random stabilizer state, and applying the unitary transformation $V_x^\dagger$. Then, we prepare the $k$ auxiliary qubits in the state $\ket{\mu} = \sum_{j \in\{0,1\}^k} (\sqrt{\mu})^{k-s(j)} (\sqrt{1-\mu})^{s(j)} (-1)^{\ell(j)} \ket{j}$. This can be achieved by a circuit composed of $k$ 2-qubit gates.

We now apply the inverse of the circuit shown on Figure \ref{fig:circuit}, producing the state 
\begin{equation}
\sum_a \alpha_a g(p_a,\mu) \ket{q_a^1}\otimes\ket{0_k} + \ldots
\end{equation}
where the ellipsis represents terms outside the image of $Q\otimes \kb{0_k}{0_k}$. The filter function is defined by
\begin{align*}
g(p,\mu) &=  \sum_{\substack{ j\in\{0,1\}^h \\  j_h = 1}} \sqrt{\mu^{h-s(j)}(1-\mu)^{s(j)}p^{h-s(j)}(1-p)^{s(j)}} \\
&=  \sum_{\ell=0}^{(k-1)/2} \binom{k}{2\ell} \sqrt{\mu^{k-2\ell}(1-\mu)^{2\ell}p^{k-2\ell}(1-p)^{2\ell}}\\
&\approx  \frac{1}{2} \exp\left\{-\frac{(p_i-\mu)^2}{2\epsilon^2}\right\}.
\end{align*}
In the last step, we have approximated the binomial distributions by normal distributions. At this point, we can use Grover's algorithm to amplify the image of $Q\otimes \kb{0_k}{0_k}$, and the rest of the analysis proceeds as above.

\end{document}